\documentclass[10pt]{article}
\usepackage{amsmath}
\usepackage{graphicx}
\usepackage{amssymb}
\usepackage[utf8]{inputenc}
\usepackage{csquotes}
\usepackage{dcolumn}
\usepackage{bm}
\usepackage{amsmath}
\usepackage{subcaption}
\usepackage[margin=1in]{geometry}
\usepackage{float}
\usepackage[T1]{fontenc} 
\usepackage[linktoc=all]{hyperref}
\hypersetup{
	colorlinks=true,
	linkcolor=blue,
	citecolor=red}

\title{Dissipative $\Lambda$CDM model with causal sign-switching bulk viscous pressure.}

\author{Vishnu A Pai$^{1,*}$, Sarath N$^{2,\dagger}$ and Titus K Mathew$^{1,\ddagger}$ \\ \small \textit{$^1$Department of Physics, CUSAT, Kalamasserry, Kochi 682022, Kerala, India} \\ \small
	\textit{$^2$Indian Institute of Technology, Kanpur 208016, India} \\ \small
	vishnuajithj@cusat.ac.in$^{*}$, sarathn@iitk.ac.in$^{\dagger}$, titus@cusat.ac.in$^{\ddagger}$ and}

\date{\today}

\begin{document}
	
	\maketitle
\begin{abstract}
Extending the standard $\Lambda$CDM model by considering dissipative effects within a causal viscous framework, and obtaining an analytical solution for the Hubble parameter remains a challenge in the literature. In this work, we resolve this dilemma by deriving a complete and original solution for the Hubble parameter by introducing a novel form for the bulk viscous coefficient associated with bulk viscous dark matter (vDM). A thorough analysis of the model is conducted by deriving theoretical constraints on the parameters and comparing the model with the latest observational data sets. Intriguingly, we find that the model predicts a sign-switching bulk viscous pressure, which facilitates both the early decelerated expansion and the late accelerated expansion of the universe. Also, the redshift at which the viscous pressure switches sign is found to be strongly correlated with the relaxation time parameter of the viscous fluid. Thermodynamic analysis revealed that, the model satisfies both the covariant and generalized second law of thermodynamics as well as the convexity condition for entropy. Additionally, we reconstructed the model by unifying viscous dark matter and dark energy into a single unified dark matter (UDM) component, and found that this unified model predicts identical dynamical evolution for the Universe, while satisfying the necessary near-equilibrium condition throughout that evolution (both in early and late phases).
\end{abstract}

\section{Introduction}\label{1}

Standard $\Lambda$CDM model in cosmology was proposed to explain the observed recent accelerated expansion of the universe. It is derived by reintroducing the cosmological constant ($\Lambda$) into Einstein's field equations and modeling dark matter as an ideal pressure-less fluid known as Cold Dark Matter (CDM). In spite of its agreement with with observational data \cite{aghanim2020planck,Scolnic_2018}, $\Lambda$CDM model faces significant challenges, such as the cosmic coincidence problem \cite{aghanim2020planck,Scolnic_2018}, the cosmological constant problem \cite{RevModPhys.61.1, Carroll:2000fy}, Hubble tension \cite{Hu:2023jqc}, S8 tension \cite{verde2019tensions,PhysRevD.107.083527} etc.. These limitations have led researchers to explore extended versions of the standard model by incorporating new physics, or derive entirely different cosmological models via alternative cosmological frameworks. Such modifications often involve altering the gravitational part of the Einstein-Hilbert action to induce changes in the dynamics of the universe \cite{Capozziello:2002rd,Dvali:2000hr, PhysRevD.75.084031,Sotiriou:2008rp} or adjusting the matter stress-energy tensor, either by introducing exotic varying dark energy components \cite{Dou:2010mz, Gagnon:2011id}, or  by modifying the properties of already hypothesized dark matter sector \cite{Mohan:2017poq}.

The perfect fluid assumption considered in the $\Lambda$CDM model could only represent an approximation to reality as the fluids found in nature has some amount of dissipation associated with them. Hence, an obvious extension of the standard $\Lambda$CDM model is to consider dissipative effects in the dark matter sector. In literature, majority of such extension of the standard model are done by utilising the Eckart's formalism to model bulk viscosity \cite{Hu:2020xus, Sasidharan:2023ndf,daSilva:2018ehn}. However, being a first order theory, this formalism violates causality \cite{ISRAEL1979341, Coley:1995dpj} and provides only unstable final equilibrium states \cite{PhysRevD.31.725}. This motivates authors to consider Israel-Stewart theory (FIS theory) for incorporating bulk viscosity in the dark sector, as it may provide causal \& stable solutions. However, the incorporation of cosmological constant in such cases, introduces non-linearity into the differential equations governing the dynamics \cite{Cruz:2018yrr}. This occurrence of non-linearity completely negates the possibility of obtaining an analytical solution for the Hubble parameter, which in turn prohibits one's efforts to understand the dynamics of the Universe in a transparent way. Furthermore, the numerical analysis carried out using those differential equations, in the context of dissipative $\Lambda$CDM model, predict abnormally large value for the bulk viscous coefficient, which even turns negative for a particular case \cite{Cruz:2018yrr}. It is therefore desirable to find an extended version of the $\Lambda$CDM model, which not only accounts for bulk dissipation in the dark matter sector, but also provides analytical solution for Hubble parameter. In the present work we aim to find such an extension of the standard model. This will then enable us to better constrain the dissipative parameters in the model and gain a better understanding about the dark sector and associated physical processes. 
	
One of the crucial ingredients in dissipative cosmological models is the form of bulk viscous coefficient. This choice not only determines the magnitude and evolution of the bulk viscous pressure but also governs the dynamics of both the early and the late Universe. Hence, the feasibility of a dissipative cosmological model mainly depends on the suitable choice of this very coefficient. However, owing to the lack of knowledge about the exact origin of bulk viscous dissipation in the cosmological context, the bulk viscosity coefficient is often chosen phenomenologically. In most cases, this coefficient is assumed to depend on factors such as the energy density of the cosmic fluid, the expansion rate of the Universe, or both. However, one must note that such choices are discretionary rather than a rule. In this work, we propose that since bulk viscosity is a transport phenomenon linked to entropic flux and heat exchange within a system, the coefficient of bulk viscosity should depend on the enthalpy density \cite{PhysRevD.91.043532} of the dissipative fluid rather than its energy density. Building on this idea, we postulate a bulk viscosity coefficient that depends on enthalpy density rather than just its energy density. This approach introduces an essential equation of state dependence to the bulk viscosity coefficient, a feature absent in previous models discussed in the literature.	

In this article, we adopt Truncated Israel-Stewart (TIS) theory to model bulk viscosity \cite{ISRAEL1979341, PhysRevD.48.1597}, primarily because it is a relatively simple second-order framework compared to the Full Israel-Stewart (FIS) formalism. Also, like FIS theory, TIS formalism is also causal and stable in the linear regime \cite{PhysRevD.107.114028, HISCOCK1983466, OLSON199018}, and is sometimes preferred over FIS theory. Moreover, there is significant evidence favoring the TIS theory over the FIS formalism  \cite{PhysRevLett.122.221602, 2021CQGra..38n5016J}. Also, some studies suggest that TIS can either be seen as a more constrained version of FIS or as a distinct phenomenological extension \cite{Piattella:2011bs}. Hence, by using the TIS formalism to incorporate viscosity, along with a novel "enthalpy density"-dependent bulk viscous coefficient, we demonstrate that it is possible to obtain an analytical solution for the Hubble parameter even in the presence of a cosmological constant and other independently evolving cosmic components.

The article is organized as follows: In Sec. \ref{2}, we will introduce bulk viscosity in the standard $\Lambda$CDM model and formulate the $\Lambda$vDM model of the Universe. In Sec. \ref{3}, we impose theoretical and observational constraints on the model parameters and estimate their best-fit values. Following this, we investigate the evolution of relevant cosmological observables in Sec. \ref{4} and conduct a comprehensive thermodynamic analysis of the model in Sec. \ref{5}. Subsequently, in Sec. \ref{6}, we develop a Unified Dark Matter (UDM) interpretation for this model and hence show that, under the said UDM interpretation, the present model satisfies the \enquote{near equilibrium condition} defined in TIS theory in Sec. \ref{7}. Finally, in Sec. \ref{8}, we conclude the study by highlighting the critical results from this investigation.

\section{FLRW Universe with causal viscous matter and cosmological constant} \label{2}

In Einstein's gravity, the Friedmann equations describing the evolution of a flat, homogeneous and isotropic Universe with bulk viscous dark matter (vDM) and cosmological constant ($\Lambda$), as the cosmic components takes the form,
\begin{eqnarray}1
	&3H^2=\rho_{m} +\Lambda \label{F1} \\
	&2\dot{H}+3H^2=-\left(p_{m}+\Pi-\Lambda\right).\label{F2}
\end{eqnarray}
Here, $H=\dot{a}/a$ represents the Hubble parameter of the universe and an over dot signifies derivative with respect to cosmic time. Also, $\rho_m$ denotes energy density, $p_m$ represents kinetic pressure and $\Pi$ corresponds to the bulk viscous pressure of vDM component. Conservation equation associated with vDM component is then given by,
\begin{equation}\label{vdmcon}
	\dot{\rho}_{m}+3H\left(\rho_{m}+p_{m}+\Pi\right)=0
\end{equation}
In order to obtain the solution to equation (\ref{vdmcon}) one must first specify the equation of state of vDM component as well as the evolution of bulk viscous pressure. Following the conventional approach we consider an equilibrium barotropic pressure of the form $p_{m}=\omega_0 \rho_{m},$ where $\omega_0$ is the constant barotropic equation of state parameter, and model bulk viscosity using TIS theory. The evolution of the bulk viscous pressure of vDM is then,
\begin{equation}\label{TIS}
	\Pi + \tau \dot{\Pi}= -3\zeta H.
\end{equation}
Here, $\tau$ is the relaxation time and $\zeta$ is the bulk viscous coefficient. According to this equation, gaining an exact understanding about the evolution of bulk viscous pressure necessitates one to know the exact dependence of $\tau$ \& $\zeta$ on the local cosmological/thermodynamic variables. By perturbation the equation (\ref{TIS}) by demanding causality and stability of solutions, Marteens derived a relation connecting these variables as \cite{maartens1996causal}, 
\begin{equation}\label{tau}
	\tau=\frac{\zeta}{\epsilon_0 \left(\rho_{m}+p_{m}\right)\left(1-c^2_{s}\right)}.
\end{equation}
where,  $c_{s}^2=\partial p_{m}/\partial \rho_{m}$, is the adiabatic speed of sound in the medium and $\epsilon_0$ is a constant free parameter, which characterizes the contrast between the speed of sound and speed of the propagation of viscous perturbations through the fluid, and it have value within the domain $ (0,1]$.
Then, by using (\ref{tau}) in (\ref{TIS}), and taking into account the barotropic nature of equilibrium pressure, one obtains the evolution equation for bulk viscous pressure as,
\begin{equation}\label{TIS2}
	\Pi + \frac{\zeta}{\epsilon_0\left(1-\omega_{0}^2\right)\rho_{m}} \; \dot{\Pi}= -3\zeta H.
\end{equation}
Finally, to close the set of equations, we must suitably postulate the form of bulk viscous coefficient. Even though literature consider the ansatz, $\zeta \propto \sqrt{\rho}$. But still there is room for better choice for $\zeta$, as one is not yet entirely sure what the exact form of $\zeta$ is. In the present study, we postulate a novel form for the bulk viscous coefficient based on two assumptions. Firstly, since bulk viscosity is a transport phenomenon directly related to entropy flux \cite{PhysRevD.53.5483}, which in turn depends on total heat content in the system, we assume that bulk viscous coefficient depends on enthalpy density of the fluid, $h=\rho_{m}+p_{m}$, rather than just on its energy density\footnote{Note that for a thermodynamic system, the total heat content is represented by the enthalpy of the system and not its internal energy density.}. This formulation basically provides an \enquote{equation of state} dependence to the bulk viscous coefficient $\zeta$, which is absent in the conventional considerations\footnote{Note that a similar enthalpy dependent contribution to bulk viscosity, namely the dynamic velocity associated with bulk viscous matter, is also postulated in \cite{PhysRevD.91.043532}}. Secondly, one also expect that the bulk viscous coefficient to depend on the expansion rate of the fluid, that is the Hubble parameter of the Universe. With these two basic considerations, the simplest ansatz that we propose for the bulk viscous coefficient $\zeta$ (which has the dimension as $\zeta$) is of the form;
\begin{equation}\label{zeta}
	\zeta \propto \frac{\rho_{m}+p_{m}}{H} \implies \zeta = \zeta_{0}\left[\frac{\rho_{m}+p_{m}}{H}\right].
\end{equation} 
Here, $\zeta_{0}$ is a dimensionless proportionality constant.

An interesting feature of the above choice is that, in the case of CDM, which has pressure $p_{m}=0$, we get $\zeta=\zeta_{0}\rho_{m}/H$, which coincidentally aligns with an earlier proposal \cite{Gomez:2022qcu}. Also, for dark energy dominated epoch, with equation of state, $p_{de}=-\rho_{de}$, equation (\ref{zeta}) can be analogously recast by replacing $\rho_m$ and $p_m$ by $\rho_{de}$ and $p_{de}$ respectively, which then implies that $\zeta=0$. Which means, the de-Sitter epoch cannot be dissipative. This is exactly what one might expect, because a de-Sitter phase corresponds to a stable equilibrium state of the universe, and therefore any out-of-equilibrium contributions such as $\Pi$ should vanish.

\subsection{Evolution of Hubble parameter}

Now will determine the analytical solution for Hubble parameter for the viscous universe by using the equation for the viscous coefficient given in equation (\ref{zeta}). We will start by substituting $p_{m}=\omega_0 \rho_{m}$ in the conservation equation (\ref{vdmcon}) and then apply a change of variable from `$t$' to `$x=\ln(a)$'. Henceforth, we obtain the evolution equation for bulk viscous pressure as,
\begin{equation}\label{pi}
	\Pi=-\left[ \frac{\rho_{m}^{\prime}}{3}+(1+\omega_0)\rho_{m} \right].
\end{equation}
Here, `$prime$' denotes a derivative with respect to `$x$'. On further differentiating (\ref{pi}) with respect to $x$, and combining it with (\ref{TIS2}) \& (\ref{zeta}), we obtain the second order linear differential equation for $\rho_m$ as,
\begin{equation}\label{DE}
	\alpha\rho_{m}^{\prime \prime}+\beta\rho_{m}^{\prime}+\gamma\rho_{m}=0,
\end{equation}
Here the coefficients are of the form,
\begin{eqnarray}
	\alpha=&\;\zeta_{0}\left[\epsilon_0(1-\omega_0)\right]^{-1}, \label{alpha}\\
	\beta=&1+3\alpha (1+\omega_0), \label{beta}\\
	\gamma=&\;\;3(1+\omega_0)(1-3\zeta_0).\label{gamma}
\end{eqnarray}
On solving this equation, we get the evolution of vDM density as,
\begin{equation}\label{rhosol}
	\rho_{m}=\mathcal{C}_1 a^{-\mathcal{N}_{1}}+ \mathcal{C}_2 a^{-\mathcal{N}_{2}}
\end{equation}
Here, $\mathcal{N}_{1}$ and $\mathcal{N}_{2}$ are constants give by,
\begin{equation}
\mathcal{N}_{1}=n_1-n_2 \quad \quad  \textbf{\&} \quad \quad \mathcal{N}_{2}=n_1+n_2, 
\end{equation}
 with $n_1$ and $n_2$ defined as,
\begin{equation}\label{n}
	n_1=\frac{\beta}{2\alpha} \quad \quad \textbf{\&} \quad \quad 
	n_2=\frac{\sqrt{\beta^2-4 \alpha \gamma}}{2\alpha}.
\end{equation}
While $\mathcal{C}_1$ and $\mathcal{C}_2$ are integration constants, the values of which are to be determined by applying suitable boundary conditions. Imposing the present condition\footnote{Here, $\Pi_0$, $H_0$ and $a_0$ are the present values of bulk viscous pressure, the Hubble parameter and scale factor respectively.}, $H \to H_0$ and $\Pi \to \Pi_0$, as $a \to a_0=1$, in equations (\ref{vdmcon}) and (\ref{rhosol}), we obtain the two independent equations connecting these coefficients which upon solving gives,
\begin{align}
	&\tilde{\mathcal{C}}_1=\left[\frac{\mathcal{N}_{2} -3(1+\omega_0)}{2n_2} \right]\left(1-\Omega^0_{\Lambda}\right)-\frac{3\Omega^0_{\Pi}}{2n_2},\label{tc1}\\
	&\tilde{\mathcal{C}}_2=1-\Omega^0_{\Lambda}-\tilde{\mathcal{C}}_1.\label{tc2}
\end{align}
 Here, $\tilde{\mathcal{C}}_1=\mathcal{C}_1/3H_0^2$, $\tilde{\mathcal{C}}_2=\mathcal{C}_2/3H_0^2$, $\Omega^0_{\Lambda}=\Lambda/3H^2_0$, 
 $\Omega^{0}_{\Pi}=\Pi_0/3H^2_0.$ Now following equation (\ref{F1}), we can write the equation for the Hubble parameter as, 
	\begin{equation}\label{Hsol}
		\mathbf{H=H_0\sqrt{\Omega^0_{\Lambda}+ \tilde{\mathcal{C}}_1 a^{-\mathcal{N}_{1}}+
		\left[1-\Omega^0_{\Lambda}-\tilde{\mathcal{C}}_1\right] a^{-\mathcal{N}_{2}}   }}
	\end{equation}
One of the intriguing possibility observed in this model is that, if $n_1 \approx n_2$, (which can  happen if $\zeta\approx1/3$ or $4\alpha \gamma/\beta^2\ll1$), then the above solution reduces to,
\begin{equation}\label{rHsol}
	H\approxeq H_0\sqrt{\tilde{\Omega}^0_{\Lambda} +(1-\tilde{\Omega}^0_{\Lambda}) a^{-2n_1}} \quad ; \left\{ \tilde{\Omega}^0_{\Lambda}=\Omega^0_{\Lambda}+\tilde{\mathcal{C}}_1\right.
\end{equation}
and if $n_1\approx 3/2$, model becomes equivalent to $\Lambda$CDM model, but with an effective cosmological constant $ \tilde{\Omega}^0_{\Lambda}$.
 
\section{Constraints on model parameters}\label{3}

Introducing bulk viscous effect in the standard $\Lambda$CDM model has introduced 4 additional free parameters in the model, all of which are related to vDM component, i.e, [$\Omega_{\Pi}^0, \zeta_0, \epsilon_0$, $\omega_o$]. Determining their exact value is essential for gaining a robust understanding about the dark sector. Hence, in this section, we will constrain the parametric space of these new variables based on certain theoretical requirements, and then determine their best fit values by comparing the model with observational data sets. 

\subsection{Theoretical Constraints}\label{3.1}

In this subsection we will constrain the parameters by demanding that this model (i) predicts a late-accelerated expansion of the Universe, (ii) obeys the covariant second law of thermodynamics (CSLT), (iii) satisfies the weak energy condition ($\rho_{m}\geq0$), (iv) maintains causality and (v) satisfies null-energy condition ($\rho_{m}+p_{m}\geq 0$) thereby avoiding phantom behaviors. 

Firstly, notice that the only acceptable value of $\omega_0$ is within the interval $[0,1)$. This is because, for any value of $\omega_0>1$ in (\ref{tau}) the relaxation time becomes negative, which is unacceptable, and for any $\omega_0<0$ we will have a negative value for squared speed of sound ($c_s^2=\omega_{0}$) in the medium. Secondly, to satisfy CSLT\footnote{In TIS theory, for satisfying the second law of thermodynamics requires $S^{\mu}_{\;\; ;\mu}=\Pi^2/\left(\zeta T\right)\geq0.$} as given in equation (2.5) in \cite{maartens1996causal}, it is essential to constrain the parameters such that the bulk viscous coefficient is always greater than zero, and from (\ref{zeta}) this implies the condition $\zeta_0>0$. Thirdly, for satisfying weak-energy condition (given as $\rho_{m}\geq0$) throughout the evolution of the Universe, it is essential to constrain the parameters such that, $\tilde{\mathcal{C}}_1\geq0$ and $\tilde{\mathcal{C}}_2\geq0$. Finally, for maintaining causality, the values of $\epsilon_{0}$ must be restricted within the domain\footnote{By carrying out perturbation analysis of a similar yet different model, a much more restrictive prior for relaxation time parameter was found in \cite{Piattella:2011bs} ($\epsilon_{0} \in [10^{-11},10^{-8}]$). However, the said prior is model dependent and hence we will not be using it in our analysis.} $\epsilon_{0}\in (0,1]$.  

To safeguard the model from phantom behaviors and violations in null energy condition, we must develop further constraints on model parameters by analyzing the behavior of effective equation of state ($\omega_{e}$) and deceleration parameter ($q$). By combining (\ref{F1}), (\ref{F2}) and (\ref{Hsol}), we obtain the effective equation of state and the deceleration parameter as,
\begin{align}
    \omega_{e}=\frac{ \tilde{\mathcal{C}}_1 \mathcal{N}_{1}a^{-\mathcal{N}_{1}}+ \left[1-\Omega^0_{\Lambda}-\tilde{\mathcal{C}}_1\right]  \mathcal{N}_{2} a^{-\mathcal{N}_{2}} }{3\left[ \Omega^0_{\Lambda} +  \tilde{\mathcal{C}}_1 a^{-\mathcal{N}_{1}}+  \left[1-\Omega^0_{\Lambda}-\tilde{\mathcal{C}}_1\right]  a^{-\mathcal{N}_{2}} \right]}-1&\label{weff}\\
    q=\frac{  \tilde{\mathcal{C}}_1 \mathcal{N}_{1}a^{-\mathcal{N}_{1}}+ \left[1-\Omega^0_{\Lambda}-\tilde{\mathcal{C}}_1\right]  \mathcal{N}_{2} a^{-\mathcal{N}_{2}} }{2\left[ \Omega^0_{\Lambda} +   \tilde{\mathcal{C}}_1 a^{-\mathcal{N}_{1}}+ \left[1-\Omega^0_{\Lambda}-\tilde{\mathcal{C}}_1\right] a^{-\mathcal{N}_{2}}  \right]}-1& \label{q}
\end{align}

From (\ref{Hsol}), it is clear that, due to the presence of $\Lambda$, this model can surely predict an accelerated expansion for the universe. However, according to (\ref{q}), for this model to have an initial deceleration era, one must at least have either, $\mathcal{N}_{1}>2$ or $\mathcal{N}_{2}>2$. From (\ref{weff}) and (\ref{q}), it is also clear that, the model shows far future phantom behaviors for the universe if $\mathcal{N}_{1}<0$ or $\mathcal{N}_{2}<0$. Hence, for the model to be devoid of phantomness, we must constrain the parameters such that, $\mathcal{N}_{1}\geq0$ and $\mathcal{N}_{2}\geq0$. Subsequently, using relations (\ref{alpha}), (\ref{beta}),  (\ref{gamma}), (\ref{n}) and considering the prior values of $\omega_0$, $\epsilon_0$ and $\zeta_0$ stated in the earlier paragraph, we notice that both $n_1$ and $n_2$ cannot be negative. Hence, combining this knowledge with the inequality $\mathcal{N}_{1}\geq0$, we obtain a single refined constraint on $n_1$ and $n_2$ as, $0<n_2\leq n_1$. This single constraint will then ensure that the model is free from any phantom behavior. Additionally, notice that this refined constraint in which we have both $n_1>0$ and $n_2>0$, also satisfies the requirement for having an early decelerated expansion (i.e, $n_1+n_2>2$).

Finally, the constraint $0<n_2\leq n_1$, together with equation (\ref{n}), implies the condition that $\gamma\geq0$. And from (\ref{gamma}), this is achievable only if $\zeta_0\leq 1/3$. Thus, the exact value of bulk viscous coefficient $\zeta_{0}$ is crucial in determining whether the model predicts a phantom era in the far future. Specifically, for values of $\zeta_0$ within the range$ [0,1/3],$ the condition $ 0 < n_2 < n_1 $ is met, resulting in a model devoid of phantom behavior and predicting a universe that asymptotically approaches a de Sitter phase in its late epochs. while for $\zeta_0 > 1/3$ the model exhibits phantom behavior in the distant future. In this study, since we require vDM to satisfy both weak energy and null energy conditions, we will constrain the bulk viscous coefficient to the range $\zeta_0 \in [0,1/3].$ 

\subsection{Observational constraints}\label{3.2}

We will now compare the model with observational datasets by performing a $\chi^2$ analysis, whilst incorporating the theoretical constraints and priors established in the previous section. For this analysis, we will use the latest Observational Hubble Data (OHD), Pantheon+ data, Baryon Acoustic Oscillation (BAO) data, and the shift parameter observed in the Cosmic Microwave Background (CMB).

\begin{table}
	\renewcommand{\arraystretch}{1.5}
	\centering
	\resizebox{\textwidth}{!}{\begin{tabular}{|c|c|c|c|c|c|c|c|c|}\hline
			Data Combinations & $H_0$& $\Omega^0_{\Lambda}$& $\Omega^0_{\Pi}\times10^{-3}$& $\zeta_0$ & $\omega_0\times10^{-3}$ &$\epsilon_0\times10^{-3}$&$\mathcal{M}$&$\chi^2_{min}$\\ \hline
			PAN$^{+}$+OHD+BAO (D1) & $72.58^{+0.51}_{-0.49}$& $0.73^{+0.01}_{-0.01}$& $-4.96^{+3.35}_{-3.43}$ & $0.23^{+0.07}_{-0.01}$& $3.65^{+6.10}_{-2.74}$&$2.18^{+4.37}_{-1.69}$&$-19.29^{+0.01}_{-0.01}$&$1890.32$\\
			PAN$^{+}$+OHD+BAO+CMB (D2)& $71.69^{+0.40}_{-0.41}$& $0.70^{+0.01}_{-0.01}$& $-0.67^{+0.50}_{-1.06}$ & $0.23^{+0.07}_{-0.01}$& $0.34^{+0.52}_{-0.25}$&$0.24^{+0.45}_{-0.19}$&$-19.30^{+0.01}_{-0.01}$&$1893.03$\\\hline
	\end{tabular}}
	\caption{\label{tab1}Best estimated value of model parameters for different data combinations.}
\end{table}

\paragraph{OHD:} For our analysis we consider the data set \cite{Singh:2024kez, Pradhan:2023nha} which includes 76 Hubble parameter values within the redshift range of $0.07\leq z \leq 2.36$. The $\chi^2$ analysis between the model and this data set is a direct comparison using the relation,
\begin{equation}\label{chiohd}
	\chi^2_{\text{OHD}}=\sum_{i=1}^{n=76} \frac{\left[H(H_0,\Omega^0_{\Lambda},\Omega^0_{\Pi},\omega,\zeta_{0},\epsilon_0,z_{i})-H_{i}\right]^{2}}{\sigma^{2}_{H_i}}
\end{equation}
Here, $H_i$ is the value of Hubble parameter observed at a redshift $z_i$ with $\sigma^{2}_{H_i}$ variance. 

\paragraph{Pantheon+ data:} Recently released Pantheon+ data sample contains 1701 light curves of 1550 distinct Type Ia supernovae (SNe Ia) spanning a redshift range of $0.001\leq z \leq 2.26$  \cite{Brout:2022vxf}. Unlike OHD data set, this data sample consist of 1701 apparent magnitude data points, each observed at a particular redshift. For comparing the model with this observational data, one must first relate the Hubble parameter to apparent magnitude ($\mu$) using the expression,
\begin{equation}\label{m}
	\mu(H_0,...,z_{i})=5\log_{10}\left[\frac{d_{L}(H_0,...,z_{i})}{\text{Mpc}}\right]+25.
\end{equation}
Here, $d_{L}$ is the luminosity distance of $i^{th}$ supernovae at a redshift of $z_{i}$ which is defined as,
\begin{equation}\label{dL}
	d_{L}(H_0,...,z_{i})=c(1+z_{i}) \int_{0}^{z_{i}}\frac{dz}{H(H_0,...,z_{i})}
\end{equation}
Then, the $\chi^2$ analysis is done by minimizing the function,
\begin{equation}\label{chisneia}
	\chi^2_{\text{PAN}^{+}}=\vec{Q}^{\;T} \cdot \left(C_{\text{stat$+$sys}}\right)^{-1} \cdot \vec{Q}
\end{equation}
Here, $\vec{Q}$ is a 1701-dimensional vector defined using the relation $\vec{Q}=\mu_{i}-\mu(H_0,...,z_{i})-\text{M}$, and $C_{\text{stat$+$sys}}$ is the covariance matrix of Pantheon+ data sample. Also, $\mu_{i}$ is the observed apparent magnitude of supernovae and $\text{M}$ is the absolute magnitude of supernovae which is often treated as a nuisance parameter.
\paragraph{CMB shift parameter:} Usage of shift parameter ($\mathcal{R}$) associated with the Cosmic Microwave Background (CMB) temperature anisotropies is one of the simplest ways of constraining the free parameters in the model with data obtained from the early universe. However, one must note that the calculation of $\mathcal{R}$ from CMB is done by assuming $\Lambda$CDM model as the background \cite{Chen_2019, Elgaroy:2007bv}, and hence one must be cautious while constraining arbitrary dark energy models using $\mathcal{R}$. Nevertheless, in the present case, we expect this model to deviate only slightly from the standard model, and hence utilize shift parameter to constrain the model. The relation connecting $\mathcal{R}$ and Hubble parameter for a flat homogeneous and isotropic universe is given as,
\begin{equation}\label{R}
	\mathcal{R}=\sqrt{\Omega^0_{m}}\int_{0}^{z_r} \frac{1}{h(\Omega^0_{\Lambda},\Omega^0_{\Pi},\omega,\zeta_{0},\epsilon_0,z)}\;dz.
\end{equation}
Here, $z_r$ is the recombination redshift and $h$ represents the dimensionless Hubble parameter $h=H/H_0$. From Planck 18 analysis \cite{Chen_2019} we have $\mathcal{R}_{o}=1.7502\pm 0.0046$ at $z_r=1089.92$. The $\chi^2$ function to be minimized in this case is then,
\begin{equation}\label{chiR}
	\chi^2_{\text{CMB}}=\frac{\left(\mathcal{R}-\mathcal{R}_o\right)^2}{\sigma_{\mathcal{R}_o}^2}.
\end{equation}

\begin{figure}
	\centering
	\includegraphics[width=0.5\columnwidth]{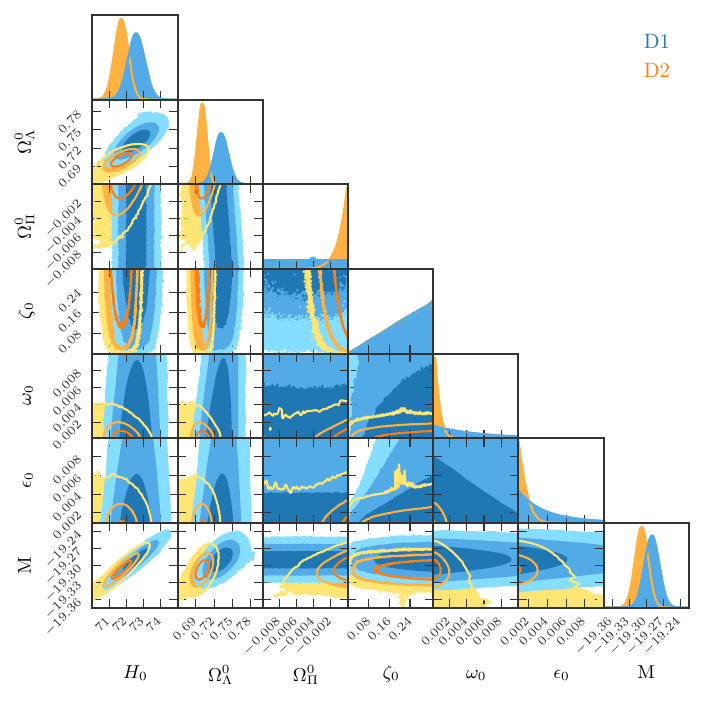}
	\caption{\label{fig1}Corner plot of 2D posterior contours with $1\sigma$ (68\%), $2\sigma$ (95\%) and $3\sigma$ (99.7\%) confidence level and 1D marginalized posteriors of all model parameters plotted using \cite{Bocquet2016} for different datasets given in Table.}
\end{figure}

\paragraph{BAO data:} BAOs corresponds to the preferred length-scale imprinted in the distribution of photons and baryons by propagating sound waves in the primordial plasma before decoupling occurred. In \cite{Ryan:2019uor}, authors found that using BAO data in association with CMB shift parameter can improve the accuracy of estimation. Hence, in the present study, we follow that approach and constrain the model using BAO data reported in \cite{BOSS:2016wmc}. This data set encompasses two distinct parameters; the transverse comoving distance $D_{M}$ (which is equivalent to line of sight comoving distance $D_{c}$ in a spatially flat universe (i.e. $D_{M}=D_{c}$)) defined as \cite{Dainotti:2022bzg},
\begin{equation}\label{Dm}
	D_{M}(z)=D_{c}(z)=\frac{c}{H_0}\int_{0}^{z} \frac{dz^{\prime}}{h(\Omega^0_{\Lambda},...,z^{\prime})}
\end{equation}
and the volume average angular diameter distance $D_{v}$ given as,
\begin{equation}\label{Dv}
	D_{V}(z)=\left[\frac{cz}{H_0}\;\frac{D_{M}^{2}(z)}{h(\Omega^0_{\Lambda},...,z)}\right]^{1/3}.
\end{equation}
The $\chi^{2}$ function to be minimized in this case is then,
\begin{equation}\label{chibao}
	\chi^2_{\text{BAO}}=\sum_{i=1}^{n=6}\frac{\left(\mathcal{D}-\mathcal{D}_o\right)^2}{\sigma_{z_i}^2}.
\end{equation}
Here, $\mathcal{D}$ and $\mathcal{D}_o$ are theoretical and observed values of BAO parameters, either $D_{M}(z_i)$ or $D_{V}(z_i)$, at redshift $z_i$, and $\sigma^{2}_{D_i}$ is the variance in those observed values. 

For comparing the present model with observations, we consider the two combinations of above mentioned data sets. The \enquote{PAN$^{+}$+OHD+BAO} data combination and the \enquote{PAN$^{+}$+OHD+BAO+CMB} data combination.  Former set constrains the parameters based on data gathered from the late phase, whereas, the latter constraints the model with CMB data also. The overall $\chi^2$ function to be minimized in each of these cases is then, $\chi^2_D=\sum_{i=1}^{N} \chi^2_i $. Here, $\chi^2_i$ represents chi-squared function of $i^{th}$ data set and $N$ denotes number of data sets involved.

\paragraph*{\textbf{Data Analysis:}} The $\chi^2$ minimization is conducted using the Markov Chain Monte Carlo (MCMC) technique, implemented via the emcee Python library \cite{Foreman_Mackey_2013}. The estimated values and variances of the model parameters are provided in Table \ref{tab1}, and the corresponding contour plot is displayed in Figure \ref{fig1}.

From the results obtained, the first noticeable feature is the consistency in the observed value of Hubble parameter despite comparing the model with different sets of observational data. That is, the model consistently predicts the present value of the Hubble parameter to be around 72 km/s/Mpc for both datasets. A careful analysis and comparison with the values reported in \cite{2021CQGra..38n5016J} reveal no significant tension between the $H_0$ values estimated from the two data sets. Therefore, by introducing a cosmological constant, we have significantly reduced the tension in the $H_0$ values found in the pure bulk viscous scenario examined in \cite{2021CQGra..38n5016J}.

Secondly, it is important to note that the equation of state parameter associated with the adiabatic speed of sound in the medium, $\omega_0$ is very close to zero. This suggests that the matter component behaves almost like viscous cold dark matter (vCDM), having a value of $\omega_0$ that is much lower than those found in \cite{2021CQGra..38n5016J}. Additionally, $\omega_0$ has an even smaller value when constrained using the D2 dataset, which includes contributions from both the early and late universe.

Finally, as seen in Fig. \ref{fig1}, the present value of the reduced bulk viscous pressure, $\Omega^0_{\Pi}$ cannot be constrained using late-time data alone (i.e., the D1 dataset). To achieve this, early-time data, such as the CMB shift parameter, is also required. Once contributions from early-time data are included, the value of $\omega^0_{\Pi}$ becomes significantly more constrained, as shown in Fig. \ref{fig1}. Therefore, we conclude that the parametric space of all model parameters related to the macroscopic properties of the fluid (such as $\epsilon_{0}$, $\omega_{0}$, $\Omega^0_{\Pi}$ and $\zeta_0$) can be more tightly constrained by incorporating additional early-universe data.

\begin{figure}
	\centering
	\includegraphics[width=.5\columnwidth]{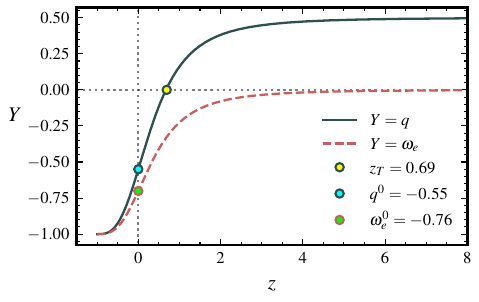}
	\caption{\label{fig2}Evolution of deceleration parameter and effective equation of state with redshift.}
\end{figure}
\section{Evolution of Cosmological Observables}\label{4}

Using the best estimated value of model parameters obtained in previous sections (D2 data set), we will now investigate the evolution of some relevant cosmological observables.

\paragraph{\textbf{Age of the universe:} } Age of the Universe is determined using the equation $Age=\int_{0}^{1}da/(aH)$. By integrating this expression by substituting Eqn. (\ref{Hsol}) and the best fit value of model parameters, we found the age predicted by the model to be approximately $13.14$ Gyrs, which is very close to $\Lambda$CDM prediction but slightly lower. Nevertheless, this estimate is in agreement with age deduced from both CMB anisotropy data \cite{ PhysRevD.74.123507} and from the oldest globular clusters \cite{Carretta_2000}. 

\paragraph{\textbf{Deceleration parameter ($q$) and Effective equation of state ($\omega_{e}$): }} Evolution of deceleration parameter and the effective equation of state, defined by relations  (\ref{q}) and (\ref{weff}) respectively, are as plotted in Fig. (\ref{fig2}). Accordingly, we learn that the model predicts the observed late-time accelerated expansion. It is to be noted that the effective equation of state parameter stabilizes around -1 in the far future, implying that universe will approach a de Sitter epoch at the asymptotic late phase of the evolution. Hence, the constraints developed in Sec. \ref{3.1} are successful in prohibiting phantom behavior from this model. Also, according to the present analysis, the transition from prior deceleration to late acceleration occurs at a redshift of $z_T=0.69$, and the current value of deceleration parameter to be $ q_0 = -0.55$, which are in conformity with observations.

\paragraph{\textbf{Dissipative variables: }}
\begin{figure*}
	\centering
	\begin{subfigure}{0.5\columnwidth}
		\centering
		\includegraphics[width=\columnwidth]{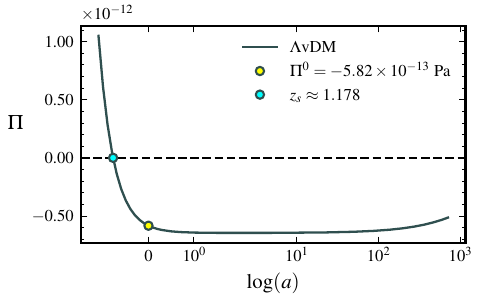}
		\caption{Bulk viscous pressure.}
		\label{fig3sub1}
	\end{subfigure}
	\begin{subfigure}{0.5\columnwidth}
		\centering
		\includegraphics[width=\columnwidth]{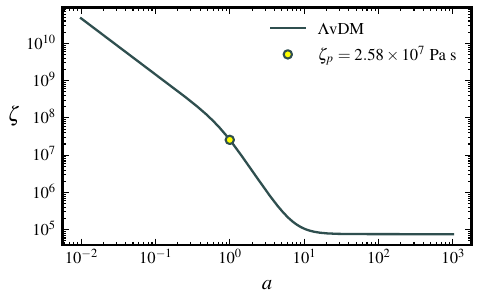}
		\caption{Coefficient of bulk viscosity.}
		\label{fig3sub2}
	\end{subfigure}
	\caption{Evolution of dissipative variables with the evolution of the Universe. In Fig. (\ref{fig3sub2}), `$\zeta_{p}$' represents the present value of bulk viscous coefficient, which is in agreement with bounds proposed in \cite{Brevik:2015jda, Sasidharan:2015ihq, e18060215}.}
	\label{fig3}
\end{figure*}

The evolution of bulk viscous coefficient and the associated dissipative pressure are obtained by plotting Eqns. (\ref{zeta}) and (\ref{pi}) respectively. Accordingly, from the plot of bulk viscous pressure, i.e, from Fig. (\ref{fig3sub1}), we see that it evolves from a positive value in the early universe, decreases gradually to zero at a redshift $z_s$ (the sign-switching redshift), and then continues to decrease further to negative values. But, at a later stage, it achieves a minimum negative value from which it then starts increasing to finally approach the value zero at asymptotic conditions. On the other hand, from Fig. (\ref{fig3sub2}), we see that the bulk viscosity coefficient remains positive throughout the evolution. This means, a positive bulk viscosity coefficient does not necessarily imply a negative bulk viscous pressure. This is an interesting behavior as compared the conventional Eckart formalism of dissipative evolution, in which a positive viscous coefficient always implies a negative bulk viscous pressure. Hence, in the present case, owing to this sign-switching behavior, bulk viscous pressure can aid both, the prior decelerated expansion, and after the sign change, the late-time accelerated expansion. Thus, in addition to the cosmological constant, bulk viscous pressure also plays a significant role in initiating the transition from the earlier decelerated phase to the current accelerated expansion. 

The asymptotic vanishing nature of viscous pressure is not fully apparent in the plot. However, it can be verified by analyzing the expression of reduced viscous pressure obtained by substituting Eqn. (\ref{rhosol}) in Eqn. (\ref{pi}) as,
\begin{equation}\label{pi2}
	\frac{\Pi}{3H_0^2}=-\left\{\frac{\tilde{\mathcal{C}}_1}{3}\left[3(1+\omega_{0})-\mathcal{N}_1\right]a^{-\mathcal{N}_1} + \frac{\tilde{\mathcal{C}}_2}{3}\left[3(1+\omega_{0})-\mathcal{N}_2\right]a^{-\mathcal{N}_2}\right\}.
\end{equation}
Using the best estimates of model parameters (D2) we obtain, $\mathcal{N}_1\approx 0.0003$ and $\mathcal{N}_2\approx 3$. Using these values in the above expression, we find that the first term on the right hand side of the above equation behaves almost like a constant, while the second term evolves as $a^{-3}$. Consequently, it can be easily concluded that, in the late stages of the evolution the first term has the dominating behavior, while in the early epochs contribution from the second term outweighs that of the first. However, it may be noted that, since the exponent $\mathcal{N}_1$ is not exactly zero, the bulk viscous pressure is not a perfect constant in the late universe, but is only a slowly decaying function of time. And in the asymptotic limit, i.e. as $z \to -1$, viscous pressure vanishes and the Universe enters de Sitter epoch driven by cosmological constant. In addition to providing the exact evolution characteristics of bulk viscous pressure, Eqn. (\ref{pi2}) also provides insights into the reason behind the sign-switching behavior of viscous pressure. Considering the constraints imposed on model parameters, i.e, $\tilde{\mathcal{C}}_1$ \&  $\tilde{\mathcal{C}}_2$ are greater than zero, and the resulting values of $\mathcal{N}_1$ and $\mathcal{N}_2,$ we obtain $3(1+\omega_{0}) - \mathcal{N}_1 >0$ and $ 3(1+\omega_{0}) - \mathcal{N}_2 < 0$. Taking account of these, the above equation then implies that, the relative viscous pressure was positive in the prior epochs and becomes negative in the later epochs.

\begin{figure}
	\centering
	\includegraphics[width=0.5\columnwidth]{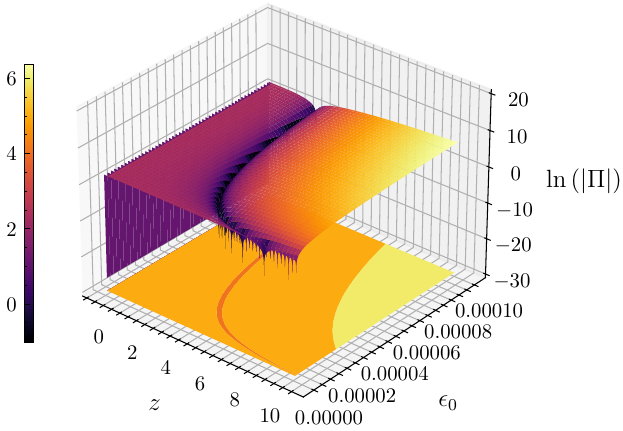}
	\caption{\label{fig4} Evolution of $\ln(|\Pi|)$ with redshift for different value of relaxation time parameter `$\epsilon_{0}$'.}
\end{figure}

From our analysis, we also found a strong correlation between the sign-switching $z_{s},$ and the relaxation time parameter `$\epsilon_{0}$'. This can be clearly visualized from Fig. (\ref{fig4}) which depicts the evolution of `$\ln(|\Pi|)$' with redshift of the Universe, for different values of `$\epsilon_{0}$'. The troughs seen in this surface plot represents different possible combinations of $(z_s,\epsilon_0),$ which corresponds to points where bulk viscous pressure undergoes sign switching. Consequently, we see that for fixed values of other parameters, a decrease in the value of relaxation time parameter causes an increase in the value of sign-switching redshift.

\begin{figure*}
	\centering
	\begin{subfigure}{0.5\columnwidth}
		\centering
		\includegraphics[width=\columnwidth]{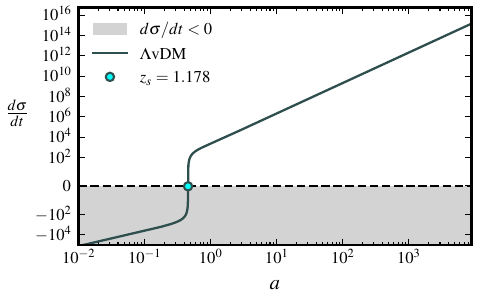}
		\caption{Specific entropy rate.}
		\label{fig5sub1}
	\end{subfigure}%
	\begin{subfigure}{0.5\columnwidth}
		\centering
		\includegraphics[width=\columnwidth]{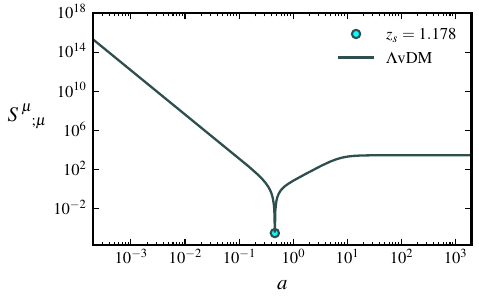}
		\caption{Entropy production rate.}
		\label{fig5sub2}
	\end{subfigure}
	\caption{Evolution of specific entropy and total entropy production rates (CSLT) with scale factor of the Universe.}
	\label{fig5}
\end{figure*}

\section{Thermodynamic analysis}\label{5}

In this section we will show that, the present model exhibits thermal evolution in conformity with the conventional laws of thermodynamics which includes; the covariant second laws of thermodynamics  (CSLT), the generalized second laws of thermodynamics (GSLT) and also the convexity condition of entropy, such that the total entropy of the system maximizes at the end epoch of the evolution. 

\subsection{Covariant Second Laws of Thermodynamics}

It is to be noted that, the evolution of total entropy production rate and specific entropy rate associated with the dissipative fluid differs in Israel-Stewart theory in contrast to the Eckart's model, where the evolution of these two factors coincide. In TIS theory the specific entropy rate evolves as \cite{PhysRevD.43.3249},
\begin{equation}\label{epro}
	\dot{\sigma}=-\frac{3\Pi H}{nT}
\end{equation}
and the total entropy production rate evolves as,
\begin{equation}\label{CSLT}
	S^{\mu}_{\;\; ;\mu}=\frac{\Pi^2}{\zeta T}.
\end{equation}
Here $n=n_0 a^{-3}$ represents the particle number density, with $n_0$ as the number density at the present epoch, and $T$ corresponds to temperature evolution of vDM which is given as \cite{PhysRevD.43.3249},
\begin{equation}\label{temp}
	T=T_{0}\;\left[\frac{\rho_{m}}{\rho^0_{m}}\right]^{\;\omega_{0}/\left(1+\omega_{0}\right)}.
\end{equation}
Here, $T_{0}$ is the present value of temperature which we set to one. By analyzing Eqns. (\ref{epro}) and (\ref{CSLT}),  by taking into account the sign-switching behavior of bulk viscous pressure, we see that the specific entropy rate evolves from negative values in the past to positive in the future (since $\dot{\sigma}\propto -\Pi$), whereas the total entropy production rate remains positive through out the evolution (as it behaves, $S^{\mu}_{\;\;;\mu}\propto\Pi^2$). Physically, this behavior can result from entropic flux leaving the comoving volume in the early universe, causing a decrease in total entropy production rate. However, since particle number density within the comoving volume doesn't change ($n^{\mu}_{\;\;;\mu}=0$), a decrease in entropy production rate in the comoving volume leads to decreasing entropy per particle in the comoving volume and hence a negative specific entropy rate. Furthermore, as universe undergoes accelerated expansion, entropic flux enters the comoving volume causing an increase in $S^{\mu}_{\;\;;\mu}$ and hence an increase in entropy per particle.
 
 We plotted the evolution of specific entropy rate and entropy production rate in Fig. (\ref{fig5sub1}) \& Fig. (\ref{fig5sub2}) respectively, by using Eqns. (\ref{zeta}), (\ref{pi}) and (\ref{Hsol}) in the above expressions. From the resulting figures we see that, as expected the specific entropy evolves form negative values, becomes zero at $z_s$ and becomes positive when viscous pressure switches its sign. On the other hand, the total entropy production rate decreases from a large positive value (which diverges at big-bang singularity), becomes zero at the sign switching redshift and then appears to be increasing towards a constant maximum value in the late phase. However, by evaluating $\lim_{z\to-1} S^{\mu}_{\;\; ;\mu} (z)$ by substituting best fit value of model parameters, one can easily verify that $S^{\mu}_{\;\; ;\mu} (z)$ does go to zero as $z \to -1$. 

\subsection{Generalized Second Law of Thermodynamics}

According to generalized second law of thermodynamics (GSLT), the total change in entropy of the Universe, which is the sum of entropy change occurring in the bulk ($S^{\prime}_{m}$) and entropy change occurring on the horizon ($S^{\prime}_{H}$), must be an increasing function in time \cite{PhysRevD.15.2738}. That is, 
\begin{equation}\label{GSLT}
	S_{H}^{\;\prime}+S_{m}^{\;\prime} \geq 0.
\end{equation}
Here, `\textit{prime}' symbolizes a derivative with respect to a suitable cosmological variable. In the present case, the most suited choice for this variable would be the scale factor of the Universe. The equation for total entropy stored in a horizon surface of area $A=4\pi\tilde{r}_{A}^2$ is $S_{H}=k_{B}A/4l_{p}^2$ \cite{Davies:1987ti}. And for a flat FLRW universe, the apparent horizon radius becomes $\tilde{r}_{A}=c/H$. Substituting this in the above equation one obtains the Horizon entropy as,
\begin{equation}\label{Sh}
	S_{H}=\left[\frac{\pi c^2 k_{B}}{l_{p}^2}\right]\frac{1}{H^{2}}.
\end{equation}
Hence, for an expanding universe, the horizon entropy is always at an increase and in its asymptotic far future, it reaches its maximum when Universe enters the end de Sitter stage.

To determine the evolution of matter entropy inside the Hubble horizon, we use the Gibbs equation,
\begin{equation}\label{gibb2}
	TdS_{m}=c^2Vd\rho_{m}+\left(c^2\rho_{m}+p_{m}+\Pi\right)dV.
\end{equation}
Using Eqns. (\ref{F1}) and (\ref{vdmcon}), along with the expression $V=(4\pi \tilde{r}_{A}^3)/3$, we simplify the above relation into,
\begin{equation}\label{sm}
	S_{m}^{\;\prime}=\frac{4\pi c^5}{3TH^3}\rho_{m}^{\;\prime}\left[1+ \frac{aH^{\;\prime}}{H}\right]=-\frac{c^5H^{\;\prime}q}{T G H^2}.
\end{equation}
Here, $q$ is the deceleration parameter of the Universe. Adding on this, the corresponding derivative of (\ref{Sh}), gives the rate of change of total entropy as,
\begin{equation}\label{ST}
	S_{T}^{\;\prime}=S_{H}^{\;\prime}+S_{m}^{\;\prime}= \frac{H^{\;\prime}}{H^{2}}\left\{\frac{c^5q}{T G}-\frac{2\pi c^2 k_{B}}{l_{p}^2H}\right\}.
\end{equation}

\begin{figure}
	\centering
	\includegraphics[width=.5\columnwidth]{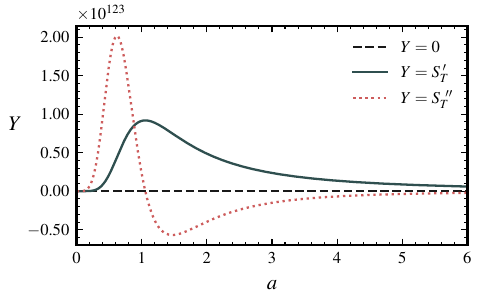}
	\caption{Evolution of total entropy and convexity condition with scale factor of the Universe.}
	\label{fig6}
\end{figure}

According to Eqn. (\ref{sm}), the matter entropy rate evolves from a positive value in the early decelerating epoch, to a negative value in the late accelerating epoch. In the far future of it's evolution, it asymptotically approaches zero from the negative side (since $H^{\prime} \to 0$ as $a \to \infty$). However, analyzing the derivative of Eqn. (\ref{Sh}), we find that Horizon entropy rate is positive throughout the evolution. And, by comparing Eqns. (\ref{Sh}) and (\ref{sm}), we can see that the magnitude of entropy change happening in the bulk is inconsequential compared to that occurring on the Hubble horizon. Hence, the total entropy change in the system mimics the behavior of horizon entropy, and is clear from Fig. (\ref{fig6}). Consequently, the model satisfies GSLT throughout the evolution.

\subsection{Convexity Condition}

In addition to satisfying the law of thermodynamics, one also expects the model to evolve towards a stable thermal equilibrium. In a closed thermodynamic system, a stable equilibrium is said to be achieved only when the entropy of that system gets maximized. The minimum requirement for this is to have the entropy evolution that satisfies the constraints $S_{T}^{\;\prime}\geq 0$ and $S_{T}^{\; \prime \prime}< 0$, at-least in the end stage of that evolution \cite{Gonzalez-Espinoza:2019vcy, callen1991thermodynamics}. The requirement $S_{T}^{\; \prime \prime}< 0$ means that the rate of increase of entropy in the universe should be decreasing, which suggest that the system is gradually approaching a stable equilibrium state. In the cosmological context, the first condition corresponds to GSLT, while the second represents the convexity condition. 

To determine whether the model complies with the convexity condition, we differentiate Eqn. (\ref{ST}) with respect to the scale factor of the Universe and plot its evolution as shown in Fig. (\ref{fig6}). We see that $S^{\;\prime \prime}_{T}$ initially increases towards a maximum positive value until the transition redshift, which is then quickly followed by a rapid decrease towards the negative side at the present epoch. From there, it evolves towards zero from the negative side in an asymptotic fashion. This evolution confirms the fact that the present cosmological model is evolving towards a stable thermodynamic equilibrium which in turn implies its agreement with the convexity condition.

\section{Reconstructing the model as a dissipative UDM model with constant adiabatic speed of sound} \label{6}

Until now we have considered $\Lambda$ and vDM as separate independent and non-interacting dark components, with the former representing dark energy and the latter signifying viscous dark matter. However, there is an interesting approach in which the cosmological constant is hypothesized to arise from equation of state of dark matter component having constant adiabatic speed of sound  \cite{PhysRevD.76.103519, PhysRevD.85.043003,Luongo:2014nld}. Here, one considers dark matter and dark energy as two faces of a single dark component called the unified dark matter (UDM). It is then interesting to check the possibility of re-interpreting the present model as a UDM model, and analyze the results based on both interpretations. Hence, in this section, we provide an alternate interpretation for the present model by considering the UDM modeling approach proposed in \cite{PhysRevD.76.103519}. In the upcoming section we will see that, adopting this UDM interpretation enables the model to satisfy the much required \enquote{near equilibrium condition} (NEC) through out the evolution of the universe, without changing the background dynamics predicted by the model. In contrast, under the \enquote{two fluid} $\Lambda$vDM interpretation, the near equilibrium condition is satisfied only in the early epoch of accelerated expansion (see the later section for more discussion).

For introducing the description of a UDM fluid, we will follow the approach made in literature \cite{PhysRevD.76.103519, PhysRevD.85.043003}. First, we define the equilibrium energy density of the unified dark fluid as $\rho_{eq}$ and it's equilibrium pressure as $p_{eq}$. With reference to the standard model, it is possible to take, $\rho_{eq}=\rho_m+\Lambda,$ (in units of $8\pi G =1$) and $p_{eq}=\omega_{e}\rho_{eq}$ \cite{PhysRevD.76.103519}. Here, $\omega_{e}(\rho_{eq})$ represents barotropic equation of state of UDM fluid. As a result the Friedmann equations are,
\begin{eqnarray}
	&3H^2=\rho_{eq} \equiv \rho_{m}+\Lambda \label{F16}\\
	&2\dot{H}+3H^2=-\omega_{e}\rho_{eq}.\label{F26}
\end{eqnarray}
Then, by assuming the adiabatic speed of sound $c^2_{s}=\partial p_{eq}/\partial \rho_{eq}$, to be a constant, we obtain,
\begin{equation}
	c_{s}^2=\omega_e +\frac{ d \omega_e}{d \rho_{eq}}\rho_{eq}=\bar{\omega}
\end{equation}
Here, $c_s^2=\bar{\omega}=constant$. One can then integrate the above equation to obtain the expression for effective equation of state of the unified fluid as,
\begin{equation}\label{weff6}
	\omega_{e}=\bar{\omega}+\frac{\mathcal{I}}{\rho_{eq}}.
\end{equation}
Here, $\mathcal{I}$ is an integration constant whose value can be determined by imposing required initial conditions. For instance, if we demand the unified dark matter to depict a cosmological constant like behavior in the asymptotic late phase, we can consider $\rho_{eq} \to \Lambda$ \& $\omega_{e}\to -1$ as $a \to \infty$, and obtain the value of integration constant as, $\mathcal{I}=-\Lambda\left(1+\bar{\omega}\right)$. If we then substitute this value of $\mathcal{I}$ in above equation we can obtain the equilibrium pressure of the unified fluid as,
\begin{equation}\label{udmeff}
	p_{eq}=\omega_{e}\rho_{eq}=\bar{\omega}\rho_{eq}-\Lambda\left(1+\bar{\omega}\right).
\end{equation}
Feeding the the general form of $\rho_{eq}$ in the right side of above equation we retain the pressure term, $p_{eq}=\bar{\omega}\rho_{m}-\Lambda$, where $\rho_{m}=\rho_{eq}-\Lambda$. This means that, the dynamics of the Universe predicted by the unified dark matter model is identical to a two-fluid $\Lambda$-WDM model of the universe, with warm dark matter (WDM) having a constant barotropic equation of state parameter `$\bar{\omega}$'. Also, note that when $\bar{\omega}\approx0$, the model becomes identical to $\Lambda$CDM case \cite{Luongo:2014nld}. 

To extend this UDM formalism in the context of the present model, we introduce bulk viscosity in the Friedmann equations obtained in previous paragraph as, 
\begin{eqnarray}
	&3H^2=\rho_{eq} \equiv \rho_{m}+\Lambda \label{F1v6}\\
	&2\dot{H}+3H^2= -\left(p_{eq}+\Pi\right) \equiv -\left(\omega_{0}\rho_{m}-\Lambda+\Pi\right). \label{F2v6}
\end{eqnarray}
Here, we have considered $\bar{\omega}=\omega_{0}$, since $\omega_{0}$ represent constant adiabatic speed of sound in the fluid medium in the presence of bulk viscosity. We will call this the \textbf{\enquote{v$\Lambda\mathbf{\bar{\omega}}$DM}} model of the Universe. However, note that, in the v$\Lambda\bar{\omega}$DM case, the viscous pressure is associated with the effective fluid as a whole, rather than with dark matter alone.

\begin{figure}
	\centering
	\includegraphics[width=.5\columnwidth]{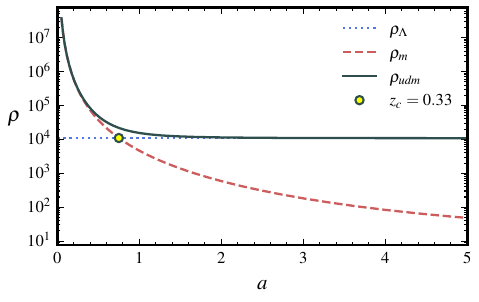}
	\caption{Evolution of energy densities in the $\Lambda$vDM and v$\Lambda\mathbf{\bar{\omega}}$DM interpretations.}
	\label{fig7}
\end{figure}%

\begin{figure*}
	\centering
	\begin{subfigure}{0.5\columnwidth}
		\centering
		\includegraphics[width=\columnwidth]{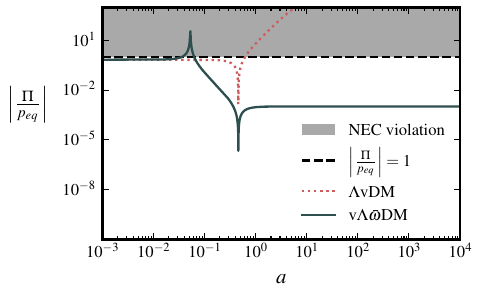}
		\caption{ }
		\label{fig8sub1}
	\end{subfigure}%
	\begin{subfigure}{0.5\columnwidth}
		\centering
		\includegraphics[width=\columnwidth]{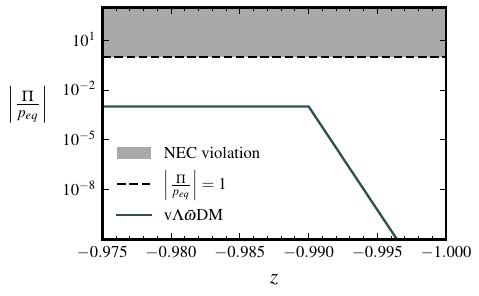}
		\caption{ }
		\label{fig8sub2}
	\end{subfigure}
	\caption{Evolution of NEC with expansion of the Universe in $\Lambda$vDM and v$\Lambda\bar{\omega}$DM interpretations for the best estimated value of model parameters. Gray regions in graph corresponds to regimes where NEC is violated.}
	\label{fig8}
\end{figure*}

In both cases, i.e, $\Lambda$vDM and v$\Lambda\mathbf{\bar{\omega}}$DM, Friedmann equations describing the dynamics of the Universe are exactly identical, as is clear from Eqns. (\ref{F1}) \& (\ref{F2}) and (\ref{F1v6}) \& (\ref{F2v6}).  Also, if we replace $\rho_{m}$ and $p_{m}$ in Eqns. (\ref{zeta}) and (\ref{tau}), with the new equilibrium variables, $\rho_{eq}$ and $p_{eq}$, we find that the relations for viscous coefficient and relaxation time are also identical in both cases. Hence, the evolution of bulk viscous pressure associated with the UDM model matches with the expressions defined in (\ref{TIS2}) and (\ref{pi}). Consequently, the three independent equations, i.e., (\ref{F1}), (\ref{vdmcon}) \& (\ref{TIS2}) that define the dynamics of the Universe, are the same in both $\Lambda$vDM and v$\Lambda\mathbf{\bar{\omega}}$DM cases. Therefore, the only difference between these two interpretations lies in the way in which one interprets the dark components, and owing to this difference, the local equilibrium variables associated with viscous fluid changes from $\rho_{m}$ \& $p_{m}$ in $\Lambda$vDM case, to $\rho_{eq}$ \& $p_{eq}$, in v$\Lambda\bar{\omega}$DM case. 

In Fig. (\ref{fig7}) we have compared the evolution of unified dark matter density in  v$\Lambda\bar{\omega}$DM ($\rho_{eq}$), with the evolution of $\rho_m$ and $\rho_{\Lambda}$ components in $\Lambda$vDM model. In the latter case, the early Universe is dominated by vDM component having a net positive pressure while the late universe is dominated by dark energy density ($\Lambda$) having a constant negative pressure. As a result, there exists a redshift ($z_{c}$) at which the energy densities of these cosmic components coincide, following which, the energy density of dark energy dominates over dark matter, thereby causing a late-accelerated expansion. While in v$\Lambda\mathbf{\bar{\omega}}$DM case, the entire dynamics of the Universe is governed only by a single cosmic component which drives both, the past deceleration as well as the late acceleration of the Universe. According to Fig. (\ref{fig7}), the energy density of this component (denoted as $\rho_{udm} = \rho_{eq}$) evolves from a singular value in the 
the early Universe towards a constant positive value in the late Universe. Consequently, the late acceleration of the Universe is achieved via the effective equation of state of UDM component, which is almost zero in the early phase and varies as $-\left(\Lambda\left(1+3\bar{\omega}\right)\right)/\rho_{eq}$ in the late universe (enabling the model to behave like of dark energy). Consequently, under UDM interpretation, the model not only satisfies NEC throughout the evolution (which is shown in the next section), but also evades the cosmic coincidence problem \cite{Piattella:2009kt}.

\section{Near Equilibrium Condition}\label{7}

Relativistic dissipative theories such as FIS and TIS theories are derived by defining a local equilibrium state for the fluid using equilibrium thermodynamic variables and then considering minute deviation from that equilibrium state. Hence, such viscous theories are valid only in near-equilibrium regimes where deviations from equilibrium states are small. In these cases, the bulk viscous pressure arises as a response to the perturbations in the dissipative medium, which drives the system back to its equilibrium state. Thus, the bulk viscous pressure is expected to die out when the fluid achieves equilibrium. These dissipative theories define viscous pressure as a deviation from the equilibrium pressure of the cosmic fluid, and as a result, the difference between the relative magnitude of bulk viscous pressure compared to that of the equilibrium pressure of the fluid acts as the measure of the deviation of the fluid from its equilibrium state. Consequently, for the fluid to remain in near-equilibrium regimes (minimal deviations from equilibrium), one must have,
\begin{equation}\label{NEC}
	\bigg| \frac{\Pi }{p_{eq}}\bigg| < 1.
\end{equation}
This constraint is often quoted in literature as the near equilibrium condition. Yet, while studying viscous-driven accelerated expansion of the Universe, one hypothesizes the validity of these dissipative theories to cases where the fluid is far from equilibrium (where NEC is violated) \cite{RMaartens_1995}. However, under such circumstances, the applicability of either TIS or FIS is questionable. Hence, having a cosmological model that abides by NEC is always preferential. In the present model, we analyze the status of NEC by adopting both $\Lambda$vDM and v$\Lambda\mathbf{\bar{\omega}}$DM interpretations. This is because, even though both models predict identical dynamics for the Universe, the NEC evolves differently in each case due to the differences in the definition of equilibrium thermodynamic variables. 

\textbf{Evolution of NEC in $\Lambda$vDM model:}  From Fig. (\ref{fig8sub1}), it is clear that the vDM component in this case evolves from a barely near-equilibrium state in the early epoch (since NEC is very close to but strictly less than one), fluctuates at sign-switching redshift (sudden dip in cure in Fig. (\ref{fig8sub1}) that arises because at $z=z_{s}$, $\Pi=0$) and turns far-from-equilibrium in the late accelerating phase. However, in the far future of its evolution, i.e, as $a\to\infty$, both $p_{eq}$ and $\Pi$ must vanish according to Eqn. (\ref{pi2}), and thus vDM eventually attains a global equilibrium state. 

\textbf{Evolution of NEC in v$\Lambda\bar{\omega}$DM model:} Evolution of NEC under UDM interpretation can be analyzed from Fig. (\ref{fig8sub2}). It is then clear that, under this interpretation the present model satisfies NEC not only in the early decelerating phase but also during the late accelerating epoch of the Universe. The NEC evolves from a near unit value (but less than one) in the early Universe, undergoes two fluctuations in the recent past and vanishes in the asymptotic far future as seen in Fig. (\ref{fig8sub2}). In this case, the fluctuation seen in Fig. (\ref{fig8sub1}), arises because of two different reasons. The initial spike in the curve where NEC grows rapidly (and goes singular) occurs due to the vanishing of equilibrium pressure, causing NEC to blow up briefly in the early Universe. Meanwhile, the second spike, which rapidly decays to zero, occurs due to the vanishing of viscous pressure at the sign-switching redshift. Furthermore, since NEC is a measure of deviation of the fluid from its equilibrium state, we infer that in this interpretation, the dissipative UDM fluid depicts an evolution from an out-of-equilibrium (but not far from equilibrium) to an equilibrium state.

\section{Results and Conclusion} \label{8}

Several authors have attempted to extend the $\Lambda$CDM model by introducing causal dissipative effects in the dark sector, and thereby propose a more generalized model of the universe. However, none of these models have succeeded in obtaining an analytical solution for the Hubble parameter in the presence of causal viscous dissipation. Such a solution is highly desirable, as it would provide stringent constraints on model parameters and deepen our understanding about the dark sector. In this article, we reconciled this long-standing challenge by obtaining an analytical solution to extended $\Lambda$CDM model in which causal viscous dissipation is accounted using Truncated Israel-Stewart (TIS) theory. Specifically, we replaced the ideal dark matter component in the standard $\Lambda$CDM model with causal bulk viscous dark matter, with a novel enthalpy-density dependent coefficient, and derived an analytical solution for the Hubble parameter. The obtained solution reveals a universe characterized by an early decelerated expansion phase followed by a late accelerated expansion phase with a pre-quintessence phase which asymptotes to a future de Sitter era, or future de Sitter epoch followed by a far-future phantom evolution.

Obtained analytical solution for the Hubble parameter predicts three possible late-time evolutionary scenarios depending on the value of the bulk viscous coefficient ($\zeta_0$), which in-turn determines the nature of vDM in the late phase. If $\zeta_{0}>1/3,$ the bulk viscous dark matter (vDM) leads to a phantom evolution. For $\zeta_{0}<1/3,$ the universe enters a quintessence phase. And notably, when $\zeta_{0}\approxeq1/3$ both the energy density and bulk viscous pressure of the dissipative matter stabilize, mimicking the behavior of a cosmological constant in the late universe. We found that, to avoid violation of null energy condition (phantom evolution) and covariant second law of thermodynamics (CSLT), $\zeta_0$ must be constrained within the range $\zeta_{0}\in (0,1/3].$

A key feature of the present model is the sign-switching behavior of the bulk viscous pressure. Initially, it starts with a large positive value, singular at the big-bang, and transitions to negative values around a specific redshift, $z_{_s}$ known as sign-switching redshift. After reaching a minimum, the pressure approaches zero asymptotically. We found that the redshift at which this sign switching occurs depends on the relaxation time parameter $\epsilon_0.$ A decrease in $\epsilon_0$ results in an increase in $z_{_s}.$ Based on theoretical constraints, we estimated the optimal value of $\epsilon_0$ to be $\epsilon_0 \approx 2.4 \times 10^{-4},$ corresponding to $z_{_s}=1.178.$

Despite the sign-switching nature of the bulk viscous pressure in this model, the coefficient of bulk viscosity remains positive throughout the evolution of the universe, hence ensuring a positive entropy production rate in the dissipative fluid and safeguarding the validity of the CSLT. This is a novel feature compared to conventional viscous models where a positive bulk viscous pressure often leads to a negative entropy production rate, thereby violating the CSLT. Interestingly, we found that during periods when the bulk viscous pressure is positive, the dissipative fluid exhibits a negative specific entropy rate in the instantaneous comoving frame. Since the model satisfies the CSLT during this era, we infer that this behavior results from a decaying entropy production rate while maintaining a constant number of particles in the comoving volume. In addition to preserving the CSLT, the model also adheres to the generalized second law of thermodynamics (GSLT) and the convexity condition, confirming that the universe evolves toward a stable equilibrium state of maximum entropy.

These findings have significant implications in viscous cosmology. For instance, in \cite{PhysRevD.59.121301} sign-switching bulk viscous pressure was considered in the acausal model to study the effect of bulk dissipation induced in the energy spectra of relic gravitons. It will be interesting to see the consequence of a more general, causal sign switching bulk viscous pressure in the context discussed in that work, which we leave as a future scope. Investigating modifications to structure formation rates in the early universe in the presence of a positive viscous pressure ia yet another intriguing possibility.

Finally, we reconsidered our model, by treating the cosmic fluids, the viscous dark matter and dark energy, as a single unified dissipative dark fluid (v$\Lambda\bar{\omega}$DM)and showed that the resultant model i.e, v$\Lambda\bar{\omega}$DM model, is dynamically similar to $\Lambda$vDM model. 
Despite the similarity in the evolutionary dynamics, the near equilibrium condition (NEC) associated with background dissipative theory evolves differently in each case. Notably, under v$\Lambda\bar{\omega}$DM interpretation, the model satisfies NEC both in the early and late Universe. While in $\Lambda$vDM 
NEC is violated in the late accelerating epoch.  In addition to this, in v$\Lambda\bar{\omega}$DM interpretation, 
evolution of NEC depicts a viscous fluid
evolving from a out of equilibrium (not far) to a stable equilibrium state. 
Consequently, we conclude that it is better to follow v$\Lambda\bar{\omega}$DM model interpretation for this model, in which the cosmological constant arises as an inherent part of the equation of state of the viscous fluid itself, rather than as an independent dark energy component.

\section*{Acknowledgments}
The authors of the manuscript are thankful to the Indian Institute of Technology Kanpur for providing the Param Sanganak high-performance computational facility for faster execution of python program. 
Vishnu A Pai is thankful to Cochin University of Science and Technology for providing Senior Research Fellowship. 
Sarath Nelleri is thankful to the Indian Institute of Technology Kanpur for providing the Institute Postdoctoral Fellowship.


\end{document}